\documentclass[12pt]{article}
\usepackage{graphicx}

\def\Re{{\cal R \mskip-4mu \lower.1ex \hbox{\it e}\,}}
\def\Im{{\cal I \mskip-5mu \lower.1ex \hbox{\it m}\,}}
\def\ie{{\it i.e.}}
\def\eg{{\it e.g.}}

\def\sub#1{_{\lower.25ex\hbox{$\scriptstyle#1$}}}
\def\tev{\,{\ifmmode\mathrm {TeV}\else TeV\fi}}
\def\gev{\,{\ifmmode\mathrm {GeV}\else GeV\fi}}
\def\mev{\,{\ifmmode\mathrm {MeV}\else MeV\fi}}
\def\mpl{\ifmmode \overline M_{Pl}\else $\overline M_{Pl}$\fi}
\def\cc{\ifmmode k/\overline M_{Pl}\else $k/\overline M_{Pl}$\fi}
\def\lpi{\ifmmode \Lambda_\pi\else $\Lambda_\pi$\fi}
\def\to{\rightarrow}

\def\subw{_{\rm w}}
\def\mh{\ifmmode m\sbl H \else $m\sbl H$\fi}
\def\mch{\ifmmode m_{H^\pm} \else $m_{H^\pm}$\fi}
\def\mt{\ifmmode m_t\else $m_t$\fi}
\def\mc{\ifmmode m_c\else $m_c$\fi}
\def\mz{\ifmmode M_Z\else $M_Z$\fi}
\def\mw{\ifmmode M_W\else $M_W$\fi}
\def\mws{\ifmmode M_W^2 \else $M_W^2$\fi}
\def\mhs{\ifmmode m_H^2 \else $m_H^2$\fi}
\def\mzs{\ifmmode M_Z^2 \else $M_Z^2$\fi}
\def\mts{\ifmmode m_t^2 \else $m_t^2$\fi}
\def\mcs{\ifmmode m_c^2 \else $m_c^2$\fi}
\def\mchs{\ifmmode m_{H^\pm}^2 \else $m_{H^\pm}^2$\fi}
\def\ztwo{\ifmmode Z_2\else $Z_2$\fi}
\def\zone{\ifmmode Z_1\else $Z_1$\fi}
\def\mtwo{\ifmmode M_2\else $M_2$\fi}
\def\mone{\ifmmode M_1\else $M_1$\fi}
\def\tb{\ifmmode \tan\beta \else $\tan\beta$\fi}
\def\xw{\ifmmode x\subw\else $x\subw$\fi}
\def\ch{\ifmmode H^\pm \else $H^\pm$\fi}
\def\lum{\ifmmode {\cal L}\else ${\cal L}$\fi}
\def\inpb{\,{\ifmmode {\mathrm {pb}}^{-1}\else ${\mathrm 
{pb}}^{-1}$\fi}}
\def\infb{\,{\ifmmode {\mathrm {fb}}^{-1}\else ${\mathrm 
{fb}}^{-1}$\fi}}
\def\epem{\ifmmode e^+e^-\else $e^+e^-$\fi}
\def\ppb{\ifmmode \bar pp\else $\bar pp$\fi}
\def\bsg{\ifmmode B\to X_s\gamma\else $B\to X_s\gamma$\fi}
\def\bsll{\ifmmode B\to X_s\ell^+\ell^-\else $B\to X_s\ell^+\ell^-$\fi}
\def\bstt{\ifmmode B\to X_s\tau^+\tau^-\else $B\to X_s\tau^+\tau^-$\fi}
\def\lamt{\ifmmode \tilde\lambda\else $\tilde\lambda$\fi}
\def\shat{\ifmmode \hat s\else $\hat s$\fi}
\def\that{\ifmmode \hat t\else $\hat t$\fi}
\def\uhat{\ifmmode \hat u\else $\hat u$\fi}

\newskip\zatskip \zatskip=0pt plus0pt minus0pt
\def\matth{\mathsurround=0pt}
\def\lsim{\mathrel{\mathpalette\atversim<}}
\def\gsim{\mathrel{\mathpalette\atversim>}}
\def\atversim#1#2{\lower0.7ex\vbox{\baselineskip\zatskip\lineskip\zatskip
  \lineskiplimit 
0pt\ialign{$\matth#1\hfil##\hfil$\crcr#2\crcr\sim\crcr}}}

\def\grtsim{\,\,\rlap{\raise 3pt\hbox{$>$}}{\lower 
3pt\hbox{$\sim$}}\,\,}
\def\lsim{\,\,\rlap{\raise 3pt\hbox{$<$}}{\lower 3pt\hbox{$\sim$}}\,\,}

\renewcommand{\thefootnote}{\fnsymbol{footnote}}

\begin{document} \begin{titlepage}
\rightline{\vbox{\halign{&#\hfil\cr
&SLAC-PUB-9596\cr
&November 2002\cr}}}
\begin{center}
\thispagestyle{empty}
\flushbottom

{\Large\bf Phenomenology on a Slice of $AdS_5
\times {\cal M}^\delta$ Spacetime}
\footnote{Work supported by the Department of
Energy, Contract DE-AC03-76SF00515}
\medskip
\end{center}

\centerline{H. Davoudiasl$^{1}$, J.L. Hewett$^{2}$, and
T.G. Rizzo$^{2}$ \footnote{e-mails:
$^a$hooman@ias.edu, $^b$hewett@slac.stanford.edu, 
$^c$rizzo@slac.stanford.edu}}
\vspace{8pt} \centerline{\it $^1$School of
Natural Sciences, Institute for Advanced Study, Princeton, NJ
08540} \vspace{8pt} \centerline{\it $^2$Stanford Linear
Accelerator Center, Stanford, CA, 94309}

\vspace*{0.7cm}

\begin{abstract}

We study the phenomenology resulting
from backgrounds of the form
$AdS_5 \times {\cal M}^\delta$, where ${\cal M}^\delta$ denotes
a generic manifold of dimension $\delta \geq 1$,
and $AdS_5$ is the slice of 5-dimensional anti-de Sitter space
which generates the
hierarchy in the Randall-Sundrum (RS) model.  The $\delta$
additional dimensions may be
required when the RS model is embedded into a more
fundamental theory.  We analyze two classes of $\delta-$dimensional
manifolds: flat and curved geometries.
In the first case, the additional flat
dimensions may accommodate localized fermions
which in turn could resolve issues, such as proton
decay and flavor, that were not addressed
in the original RS proposal.  In the latter case, 
the positive curvature of an $S^\delta$ manifold
with $\delta> 1$ can geometrically provide
the 5-dimensional warping of the RS model.  We
demonstrate the key features of these two classes of models
by presenting the background solutions,
the spectra of the Kaluza-Klein (KK) gravitons, and their
4-dimensional couplings, for the sample manifolds
$S^1/Z_2$, $S^1$, and $S^2$.  The resulting
phenomenology is distinct from that of the original RS scenario
due to the appearance of a multitude of new KK graviton states at the
weak scale with couplings that are predicted to be measurably 
non-universal within the KK tower. In addition, in the case of flat
compactifications, fermion localization can result in KK graviton
and gauge field flavor changing interactions.

\end{abstract}

\renewcommand{\thefootnote}{\arabic{footnote}} \end{titlepage}

\section{Introduction}

In the Randall-Sundrum (RS) model \cite{RS1}, the hierarchy
between the electroweak and gravitational scales is explained
in the context of a 5-dimensional (5-d) background
geometry, which is a slice of anti-de Sitter ($AdS_5$) spacetime.
Two 3-branes of equal and opposite tension sit at orbifold
fixed points at the boundaries of the $AdS_5$ slice.   We denote
this geometrical setup as $|AdS_5|$ to emphasize that it consists of a
slice of anti-de Sitter space.  The
5-d warped geometry induces a 4-d effective scale $\Lambda_\pi$
of order a TeV on one of the branes.  \lpi\ is thus
exponentially smaller than the gravitational scale, which is
given by the reduced Planck mass \mpl.
In this scenario, parameters of the 5-d theory
maintain their natural size, of order \mpl,
even though the 4-d picture has
hierarchical mass scales with $\Lambda_\pi/\mpl \sim 10^{-15}$.  The
RS proposal has distinct phenomenological signatures that are
expected to be revealed in experiments at the TeV scale, and hence
the phenomenology of
this model has been studied in detail \cite{DHRprl,RSphen}.

From a more theoretical perspective, one may view the RS model as
an effective theory whose low energy features originate from
a full theory of quantum gravity, such as string theory.  Based on
this view, and also on general grounds, one may expect that a more
complete version of this scenario must admit the presence of additional
extra dimensions compactified on a manifold ${\cal M}^\delta$ of
dimension $\delta$.
From a model building point of view, it can be
advantageous to place at least some of the Standard Model (SM)
fields in the higher dimensional space.
This possibility generally allows for new model-building techniques
to address gauge coupling unification \cite{keith1}, supersymmetry
breaking \cite{martin}, and the neutrino mass spectrum \cite{yuval}.
However, the placement of SM fields in the RS $|AdS_5|$ bulk is
problematic due to large contributions to precision electroweak 
observables
arising from the SM Kaluza-Klein (KK) states \cite{RSphen}.
Hence the presence of the additional manifold may reconcile these model
building features with the RS model.
For example, in the RS model, despite its
interesting features, the issues of proton decay and flavor do
not have a simple explanation.  However, these problems are
naturally addressed in simple geometries, such as an $S^1/Z_2$
extra dimension along which fermion fields are localized
\cite{ArkSch}.  Thus, considering an $|AdS_5 |\times S^1/Z_2$
background, for example, can in a straightforward way provide the
RS model with a geometrical explanation of proton stability and flavor,
while preserving the desirable features of its description of the
hierarchy.

Given the motivation for considering an extended $|AdS_5| \times
{\cal M}^\delta$ background, it is then interesting to determine how 
the
original RS phenomenology is modified in the presence of the
additional manifold and what
new signatures can be expected in future experiments.  Other studies
of extended RS scenarios \cite{extRS} have focused on the mechanism
for localizing gravity on a thick brane or on a singular string-like
defect, but have not examined the resulting phenomenology.
In this work, we examine the phenomenology resulting from two
classes of additional geometries; flat
and curved spaces.  We consider the manifolds $S^\delta$, with
$\delta \geq 1$, as this choice has the advantage of
simplicity and provides a representative example of each class
of geometry.   The flat case with $\delta=1$ provides a natural
mechanism for addressing proton decay and flavor as discussed
above.  For $\delta > 1$, the extra manifold
$S^\delta$  has positive curvature and we elucidate how this
curvature can serve as the origin of the warping in the 5-d RS picture.
We expect that the qualitative features of our results
obtained from $S^\delta$ are representative of the phenomenology
for fairly general choices of the manifold ${\cal M}^\delta$.

We find that the addition of the $S^\delta$ background to the RS
setup typically results in the emergence of a forest of graviton KK
resonances ocurring in between the original RS resonances.
This forest originates from the `angular' excitations on the
$S^\delta$.  In this paper, we focus mainly on the derivation of the 
graviton
KK spectrum, and study the cases ${\cal M} = S^1, S^1/Z_2$, and
$S^2$ in some detail. We address the constraints placed by data on
this picture, as well as issues related to the extraction of model
parameters from experimental data.  In the scenario where the SM
fermions are localized in the $S^1/Z_2$ manifold, we note that
flavor changing graviton interactions may arise at the tree-level,
which could in principle pose a threat to this model.
We give an estimate for such contributions 
and find that they do
not occur at a dangerous level.  We note that in $d > 5$,
gravi-scalars and gravi-vectors \cite{GiudiceHan}
also play a role in weak scale phenomenology, especially
if the SM fields reside in the $S^\delta$, but we do not consider the
modifications from these sectors here.

In the next two sections, we present the background solutions
for the flat and curved geometries, corresponding to
$\delta = 1, 2$, respectively.   In each case, we examine the resulting
spectrum of the KK gravitons and compute their couplings to 4-d fields.
We then present our
results for the expected production cross sections of the KK
gravitons at colliders, as well as the associated experimental
bounds.   Lastly, we comment on
the case where the SM fields can propagate in the $S^\delta$,
and outline some of the issues that need further study in this case.  
We
present our conclusions in section 4.

\section{The Graviton KK Spectrum and Couplings: Flat Manifolds}

For simplicity, we limit our study to the flat manifolds $S^1$
and $S^1/Z_2$; our results easily generalize to the case of
other toroidal compactifications.
We first discuss the boundary conditions that lead
to the $|AdS_5|\times S^1$ background and derive the 4-d spectrum
of the corresponding KK gravitons and their couplings.  We then
study the phenomenology of the KK graviton spectrum and discuss some
consequences of placing the SM fermions in the additional manifold.

\subsection{Formalism}

The metric for the $|AdS_5|\times S^1$ background is given by
\begin{equation}
ds^2 = e^{-2\sigma} \eta_{\mu \nu} dx^\mu dx^\nu + r_c^2 \,
d\phi^2 + R^2
\,d\theta^2,
\label{s1met}
\end{equation}
where, following the
conventions and notation of Ref.\cite{RS1}, we have $\sigma = k
r_c |\phi|$, with $k > 0$ setting the scale of the curvature of the
$AdS_5$ slice.  The `radial' (RS) dimension, $x_4$, is parameterized 
as $x_4=r_c\phi$ by the
angle $\phi \in [-\pi, \pi]$, where $r_c$ is the compactification
radius.  The two 4-branes sit at the orbifold fixed points 
$\phi=0,\pi$.
The $S^1$ is parameterized by the
angle $\theta \in [0, 2 \pi]$, and $R$ is the radius of the $S^1$.
To obtain a solution to Einstein's equation corresponding to this 
metric,
we need the inhomogeneous cosmological constant tensor and
the energy-momentum tensor for the sources on the branes 
\cite{koganmulta}.
For the cosmological constant
tensor, we choose
\begin{equation}
\Lambda^A_B =
{\rm diag}(\Lambda, \Lambda, \Lambda, \Lambda, \Lambda,\Lambda_\theta),
\label{cctensor}
\end{equation}
and the energy-momentum tensor is assumed to have the form
\begin{equation}
T^M_N = - \left\{\delta(\phi)\left(\begin{array}{ccc}
V^h \delta^\mu_\nu & 0 & 0 \cr
0 & 0 & 0 \cr
0 & 0 & V_\theta^h
\end{array}\right) + \delta (\phi - \pi)
\left(\begin{array}{ccc}
V^v \delta^\mu_\nu & 0 & 0 \cr
0 & 0 & 0 \cr
0 & 0 & V_\theta^v
\end{array}\right)\right\},
\label{emtensor}
\end{equation}
with $V$ representing the brane tensions
where the superscripts $v$ and $h$ correspond to the visible and
hidden branes, respectively.

In a fashion similar to that in Ref.\cite{RS1},
we can then solve Einstein's equations for the above
setup, and obtain the following results
\begin{equation}
V^h = - V^v = - \Lambda/k = 24 \, M_F^4 \, k,
\label{rel1}
\end{equation}
and
\begin{equation}
\Lambda = \frac{3}{5} \, \Lambda_\theta \, \, ; \, \,
V^h = \frac{3}{4} \, V^h_\theta\, \, ; \, \,
V^v = \frac{3}{4} \, V^v_\theta\, ,
\label{rel2}
\end{equation}
where $M_F$ denotes the 6-dimensional fundamental scale.
Following the conventions of Ref.\cite{RS1}, we have
\begin{equation}
\mpl^2 = \int_{-\pi}^{+\pi}r_c \, d \phi \int_0^{2 \pi} R
\, d \theta \, e^{-2 \sigma} M_F^4,
\end{equation}
from which we derive the following relation
\begin{equation}
\mpl^2 = \frac{2 \pi R}{k} \, M_F^4
[1 - e^{-2 \sigma(\pi)}].
\label{MPrel1}
\end{equation}

We now discuss the derivation of the KK spectrum.
In the rest of this work, we will limit our analysis
to the case of metric perturbations of the form
\begin{equation}
G_{\mu \nu} = e^{-2\sigma} (\eta_{\mu \nu} + \kappa \,
h_{\mu \nu}),
\label{metpert}
\end{equation}
where $\kappa = 2/M_F^{(3 + \delta)/2}$, and $M_F$ is the
fundamental scale in $(5 + \delta)$ dimensions.
Our background solutions for $S^1$ and $S^2$ follow from the
original RS convention, except for our choice of $\kappa$,
since it is assumed that the coefficient of the 5-d curvature
term is $\kappa^{-2}$.
The gauge choice for all our computations is the transverse
traceless gauge;
$\partial^\mu h_{\mu \nu} = h^\mu_\mu = 0$.  We choose the
following KK expansion for $h_{\mu\nu}$ in Eq.(\ref{metpert})
\begin{equation}
h_{\mu\nu}(x, \phi, \theta) = \sum_{n, l}
h^{(n, \, l)}_{\mu\nu}(x)
\frac{\chi^{(n,\, l)}(\phi)}{\sqrt {r_c}}
\frac{\varphi^{(l)}(\theta)}{\sqrt {R}}.
\label{gravKK}
\end{equation}
The $\theta-$dependent wavefunction is given by
\begin{equation}
\varphi^{(l)}(\theta)= e^{i l \theta}/{\sqrt {2 \pi}}
\end{equation}
in the case of $S^1$ and
\begin{equation}
\varphi^{(l)}(\theta)= \left\{ \begin{array}{ll}
                               {1/\sqrt{2\pi}}, & \quad l=0\\
                               {\cos l\theta/\sqrt\pi}, & \quad l\ne 0
                              \end{array}
                       \right.
\end{equation}
for the orbifolded case $S^1/Z_2$.
Inserting the above KK expansion in the perturbed Einstein's equations,
we find, to linear order in $\kappa$, the
following eigenvalue equation for the $(n,\, l)$ mode
with mass $m_{nl}$
\begin{equation}
-\frac{1}{r_c^2} \frac{d}{d\phi}\left(e^{-4 \sigma}
\frac{d}{d\phi} \chi^{(n,\, l)}(\phi)\right) +
e^{-4 \sigma}\left(\frac{l}{R}\right)^2 \chi^{(n,\, l)}(\phi)
= e^{-2 \sigma} m_{nl}^2 \chi^{(n,\, l)}(\phi).
\label{KKeq}
\end{equation}
This is the equation of motion in the RS model for
a bulk scalar field of 5-d mass $l/R$.
The solutions are given by \cite{GW}
\begin{equation}
\chi^{(n,\, l)}_l(\phi) = \frac{e^{2 \sigma}}{N_{nl}} \,
[J_\nu(z_{nl}) + \alpha_{nl} Y_\nu(z_{nl})],
\label{chi}
\end{equation}
where $J_\nu$ and $Y_\nu$ denote Bessel functions of order $\nu$.
We now find
\begin{equation}
\nu \equiv \sqrt {4 + \left(\frac{l}{k R}\right)^2}
\label{nu}
\end{equation}
with $z_{nl}(\phi) \equiv (m_{nl}/k) e^\sigma$.
(We will ignore terms suppressed by powers of $e^{-k r_c\pi}$
throughout our computations.)  The normalization $N_{nl}$ is then given 
by
\begin{equation}
N_{nl} = \frac{e^{k r_c \pi}}{\sqrt{k r_c}}J_\nu (x_{nl})
\sqrt {1 + \left(\frac{4 - \nu^2}{x_{nl}^2}\right)},
\label{Nq}
\end{equation}
where $x_{nl} = z_{nl}(\pi)$.  The $x_{nl}$ are solutions of the 
equation
\begin{equation}
2J_\nu (x_{nl}) + x_{nl}J_\nu^\prime(x_{nl}) = 0,
\label{xq}
\end{equation}
which yields the masses $m_{nl}$.  There are no modes for which
$l \neq 0$ and $\chi^{(n,\, l)}(\phi) = $ Constant, {\ie,} for $n=0$, 
only
the case $l=0$ is allowed.
This means that all KK graviton states have warp-factor enhanced
couplings, as in the original RS model.
The zero mode, corresponding to the massless 4-d graviton,
has the wavefunction $\chi^{(0,\, 0)} = \sqrt{k r_c}$.

Since we are interested in the situation where the SM fields
are localized on the circle, we consider, for simplicity, the case
where they are placed at $\theta=0$.  For either $S^1$ or $S^1/Z_2$,
the results for the graviton couplings derived below can be easily
generalized to the case $\theta=\theta_0$, with $\theta_0$ being
arbitrary, as will be discussed in section 2.4.2.
The 6-d graviton coupling to the energy momentum
tensor of 4-d fields localized at
$\phi = \pi$ and $\theta = 0$ is given by
\begin{equation}
{\cal L} = -\frac{1}{M_F^2} h^{\mu\nu}(x, \pi, 0)T_{\mu\nu}(x).
\label{6dcoup}
\end{equation}
Substituting the KK expansion (\ref{gravKK}) in the above,
and using the expressions for $\chi^{(n)}_l(\phi)$ and $N_n$,
yields
\begin{equation}
{\cal L} = -\frac{1}{\mpl} h_{\mu\nu}^{(0, \, 0)}(x)T^{\mu\nu}(x)
-\frac{1}{\Lambda_\pi} T^{\mu\nu}(x)
\sum_{l = -\infty}^\infty \sum_{n = 1}^\infty \xi(n, l)
h_{\mu\nu}^{(n, \, l)}(x)
\label{4dcoup}
\end{equation}
for the case of $S^1$, and
\begin{equation}
{\cal L} = -\frac{1}{\mpl} h_{\mu\nu}^{(0, \, 0)}(x)T^{\mu\nu}(x)
-\frac{1}{\Lambda_\pi} T^{\mu\nu}(x)
\sum_{l = 0}^\infty \sum_{n = 1}^\infty C_l\, \xi(n, l)
h_{\mu\nu}^{(n, \, l)}(x)
\label{4dcouporb}
\end{equation}
for the orbifolded case $S^1/Z_2$.
Here, $\Lambda_\pi \equiv e^{- k r_c \pi} \mpl$,
\begin{eqnarray}
\xi (n, l) & = & \left[1 - \left(\frac{l}
{k R x_{nl}}\right)^2\right]^{-1/2}\,, \nonumber \\
C_l & = & \left\{ \begin{array} {cl}
                    1, & \quad l=0 \\
                    \sqrt 2, & \quad l\ne 0
                   \end{array}
          \right.  \,.
\label{xi}
\end{eqnarray}
The massless zero mode graviton
couples with the 4-d strength $1/\mpl$, as required.
We note that the coupling of
various KK gravitons to the 4-d energy-momentum tensor is not
universal.  This non-universality, with natural choices
of parameters, could be of ${\cal O}(1)$ for the light KK modes,
and would be easily measured in experiment.
This is in contrast to the case of the original RS
scenario where the non-universality is
suppressed by ${\cal O}(e^{-k r_c \pi})$ for the light
modes \cite{DHRprl}.  Note that since $x_{nl}>l/kR$, the factor $\xi{(n,l)}$
is always real.

\subsection{Discussion of Model Parameters}

Here, we discuss the parameters present in this scenario and their
relation to the phenomenological features of the graviton KK spectrum.
In the original RS model{\cite {RS1}},
the two parameters can be chosen to
be the mass of the first graviton resonance (or, alternatively, 
$\Lambda_\pi$)
and the ratio $k/\mpl$. The desire that there be no new hierarchies 
suggests
that $k/\mpl \gsim 10^{-2}$ while the validity of the classical 
approximation
leads to the requirement that the 5-dimensional curvature satisfy
$|R_5| \leq \mpl^2$ which, in turn, implies
$k/\mpl \lsim 0.1$ {\cite {DHRprl,RSphen}}.
In the present scenario, there
is an additional parameter, $R$, describing the radius of the circle
$S^1$ or the orbifold $S^1/Z_2$. The
corresponding consideration of the 6-dimensional curvature, $R_6$, 
yields
the same bound as above on the ratio $k/\mpl$ while giving no 
constraint on
$R$.  However, naturalness and the wish that no new hierarchies exist
suggests that the value of the mass scale $R^{-1}$ not be very
different from $k$, $\mpl$ or
the fundamental scale $M_F$.  To be specific, we will assume that
$0.1 \lsim kR \lsim 10$ in our numerical analysis below;
this range is sufficient to demonstrate the features
of the new physics associated with the $S^1(/Z_2)$ extension
to the original RS model.
As we will see below, the value of $R$ is intimately connected with the
shape of the KK spectrum in the present scenario.

Before discussing our results, we first examine the qualitative
features of the KK spectrum in the present scenario.
For convenience, we label the states and their corresponding masses
by the values of the integers $n$ and $l$, as denoted in the
previous section.  This notation suggests that $(n,\, l)$
play a role similar to the quantum numbers
in atomic structure considerations. The states $l=0$ for any $n$ 
correspond to the KK states in the original RS construction.  
In addition, as we saw
above, for $n=0$ only $l=0$ is allowed, but for any $n \geq 1$, states 
with $l \geq 0$ exist. 

\begin{figure}[htbp]
\centerline{
\includegraphics[width=9cm,angle=90]{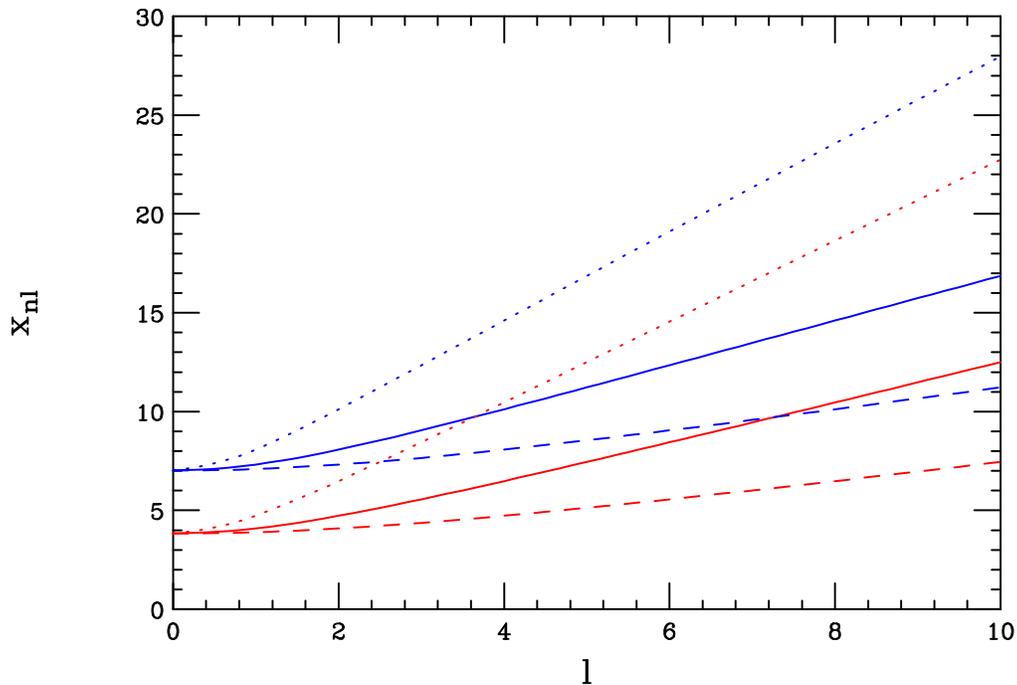}}
\vspace*{0.1cm}
\caption{The roots $x_{nl}$ for $n=1,2$ as a function of $l$.  The
dotted, solid, dashed curves correspond to $kR = 0.5,\, 1.0,\, 2.0$,
respectively.}
\label{roots}
\end{figure}

To gain an understanding of the KK spectrum
a short numerical analysis shows that we may write the roots $x_{nl}$,
which determine the KK tower masses, in the empirical form
\begin {equation}
x_{nl}^2 \simeq \Bigg[x_n^2 + {x_n l\over {kR}} +\bigg({l\over 
{kR}}\bigg)^2
\Bigg]\,,
\label{specfit}
\end{equation}
where the $x_n$ are the roots of $J_1$, \ie, $J_1(x_n)=0$, and describe
the KK spectrum in the original RS scenario \cite{DHRprl}.  This
expression is justified by the behavior of the roots as a function of
$l$ as shown in Fig. \ref{roots}.  Clearly, in this figure, only integral
values of $l$ correspond to the physical situation.
Note that the term linear in $x_n,\, l$ in the above expression
does not appear in the
conventional case of two dimensional toroidal compactifications and
has the appearance of a `cross-term' between the
contributions from the two geometrically different dimensions.
This behavior also appears in the similar situation
where the extra dimensional shape and moduli are both
considered {\cite {keith}}.

A more complete analysis of the roots $x_{nl}$ in this scenario confirms
the qualitative features of the above empirical form for
the KK tower masses. We have derived the behavior of the roots using
asymptotic forms for the Bessel functions. In the limit of large $kR$,
we find
\begin{equation}
x^2_{nl} \simeq x_n^2+ {x_n\pi\over 4}\bigg({l\over kR}\bigg)^2\,,
\end{equation}
whereas, for small values of $kR$,
\begin{equation}
x^2_{nl} \simeq \Bigg[ x_n + {\pi l\over 2kR}\Bigg]^2\,.
\end{equation}
Note the appearance of the cross term in these limits.
We find that these expressions provide a fairly accurate approximation
to the full numerical results for the roots in these two limiting cases. 

We now discuss the consequences of the
above expression; in particular, there are two limiting cases which
may hinder the observation of this scenario at colliders.
When $(kR)^{-1}$ is large in comparison to $x_n$, the
first $l$ excitation is heavy in comparison to the
$l=0$ state, \ie, $m_{n1}/m_{n0}$ is
large. This is due to the dominance of the quadratic term.
If this mass ratio is too large, it is possible that the $l\ge 1$ KK
states will be too massive to be produced directly at colliders.
Colliders such as the LHC will thus only observe the KK graviton
mass spectrum that is present in the usual RS model.
Fortunately, a short analysis indicates that this is unlikely
as long as $kR \gsim 0.05$ and hence lies within our natural
range of values.
On the otherhand, in the case where $kR$
is large, the $l$ excitations for a given $n$ will be very close
in mass to the $l=0$ mode and will have a tendency to pile up near
this state.   Here, the separations between the $l$ excitations
are tiny (yet growing with $l$), and may be too small to be observed
at colliders.  It is possible that the individual peaks would not be
isolated in the data and information about the
individual states would be lost. As we will see below, this is
not a problem for the natural range of $kR$ that we have assumed,
however, if $kR$ lies outside of this range, this issue will emerge.

\subsection{Numerical Results}

\begin{figure}[htbp]
\centerline{
\includegraphics[width=9cm,angle=90]{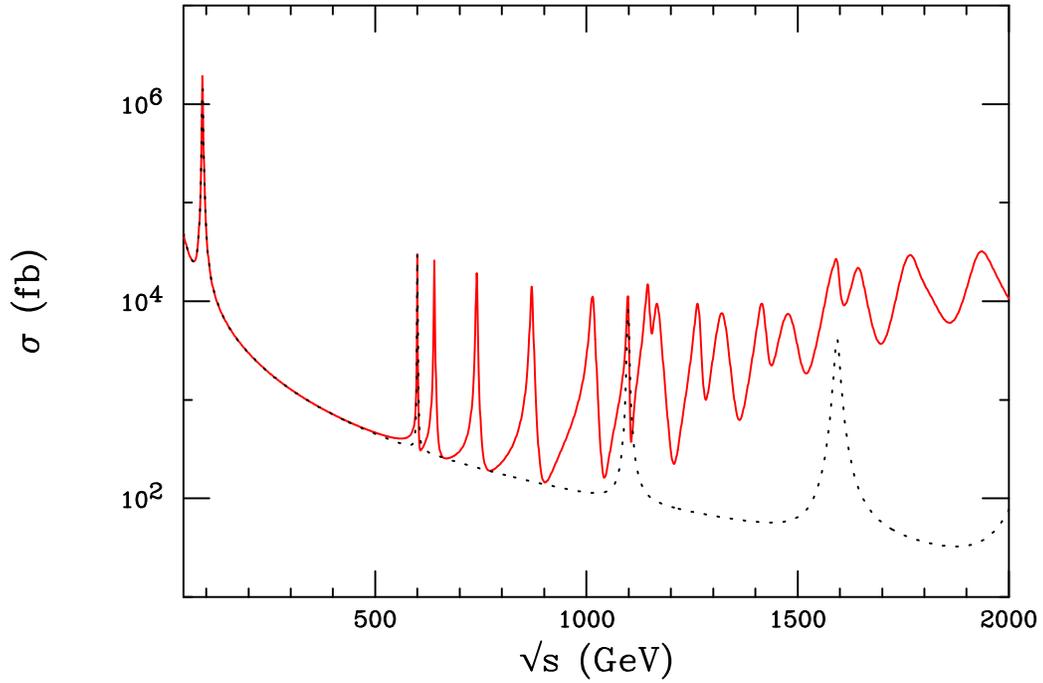}}
\vspace*{0.1cm}
\caption{The solid red curve corresponds to the cross section for 
$e^+e^- \to \mu^+\mu^-$ when the additional
dimension is orbifolded, \ie, for $S^1/Z_2$, with $m_{10}=600$ GeV, 
$k/\mpl=0.03$
and $kR=1$ being assumed. The result for the conventional RS model is 
also displayed, corresponding to the dotted curve.}
\label{fig1}
\end{figure}

We first present the numerical results for the case where the additional 
dimension is orbifolded, $S^1/Z_2$.
Fig.~\ref{fig1} displays the spectrum of KK graviton excitations
that would be observed at an $e^+e^-$ collider via the process $e^+e^- 
\to\mu^+\mu^-$.  Here, we assume $m_{10}=600$ GeV, $k/\mpl=0.03$ 
and $kR=1$; we will take this to be our reference set of parameter values
in our discussions below.   We note that the mass of the first graviton 
KK excitation is consistent
with bounds from the Tevatron data sample \cite{DHRprl}.  The
conventional KK graviton spectrum in the original RS scenario
is shown for purposes of comparison. In performing
these calculations we have made a simplifying assumption
which only influences the KK states at or above the $n=3,\, l=0$ mode.
These heavy states are allowed to decay to other KK modes \cite{hr}
(note that the value of $l$ would need to be conserved in these decays 
due to orthonormality of the wavefunctions) and are thus wider than
assumed in the present analysis which only includes decays to SM 
fields.  Such potential decay channels include not
only gravitons but also the gravi-vectors and gravi-scalars which remain 
after the KK decomposition.  The latter modes are not present in the
conventional RS model, as in that case the physical
spectrum consists solely of the graviton KK tower and the radion. 
However, for six (or
more) dimensions other physical states {\cite {GiudiceHan}}
remain after the KK decomposition; for $N$ extra
dimensions there are $N-1$ physical gravi-vector KK towers as well as
${1\over {2}} N(N-1)$ physical gravi-scalar KK towers. When the 
resonant gravitons decay to these states they will generally appear 
as missing energy at a collider. With the SM fields confined to the 
orbifold fixed points as discussed above, these additional 
fields either do not couple to those of the SM or are coupled 
sufficiently weakly that their effects can be safely ignored in the 
present discussion.  Our neglect of these additional channels
is also validated by the fact
that the lighter modes below the $n=3,\, l=0$ state are
most likely to be the ones within kinematic reach at colliders.
When $N\ge 1$, the exchange of these gravi-scalar
and gravi-vector states may contribute to the KK graviton
spectrum when the SM fields are not located at the orbifold
fixed points.

A number of features can be observed from Fig.~\ref{fig1}: 
($i$) The density of the KK
spectrum has substantially increased in comparison to the original
RS model and now resembles a `forest' of peaks, similar to the lines
that appear in an atomic spectrum. Given the assumed value $kR=1$,
both $n$ and $l$ excitations are present in the same kinematic region.
There are 19 KK states appearing in the kinematical range displayed
in this figure!
($ii$) The peaks are generally well separated except
where accidental overlaps occur, \eg, the $n=1,\, l=5$ and $n=2,\, l=0$ 
states are essentially degenerate for this choice of parameters. Some 
overlap also occurs in the 
case of the more massive states due to their larger widths.
($iii$)  Overall, the cross section at large $\sqrt s$
rises faster in the present scenario than in the usual RS model due to the 
large number of states; this may lead to early violations of unitarity and 
will be discussed further below.  $(iv)$  The properties of the lowest lying 
KK state ($n=1,\, l=0$) are not altered significantly from that found in the 
conventional RS model; its width can then be used to
extract the value of the parameter $k/\mpl$ in the usual manner {\cite 
{clica}}.

We have checked whether the rise in the cross section for large values
of $\sqrt s$ leads to early violation of unitarity.  The partial wave
unitarity bound \cite{GiudiceHan} on the cross section for $2\to 2$ scattering 
of inital and final state fermions with  helicity of 1 is given by
\begin{equation}
\sigma_U = {20\pi\over s} = 24.4 \times 10^{6} \Bigg[ 
{1{\rm TeV}\over\sqrt s}\Bigg]^2 {\rm fb} \,.
\label{unitar}
\end{equation}
We employ the criteria that the cross section be well behaved up to the 
ultra-violet scale in this theory, \ie, $\Lambda_\pi$.  For this set
of parameters we have $\Lambda_\pi\simeq 5$ TeV, and we see that
unitarity is violated as $\sqrt s$ closely approaches this value.
Either the ultra-violet effects must set in slightly below
$\Lambda_\pi$, or this set of parameters results in a mild 
violation of unitarity.  We will return to this point below when examining
the variation of the KK spectrum with the model parameters.

\begin{figure}[htbp]
\centerline{
\includegraphics[width=9cm,angle=90]{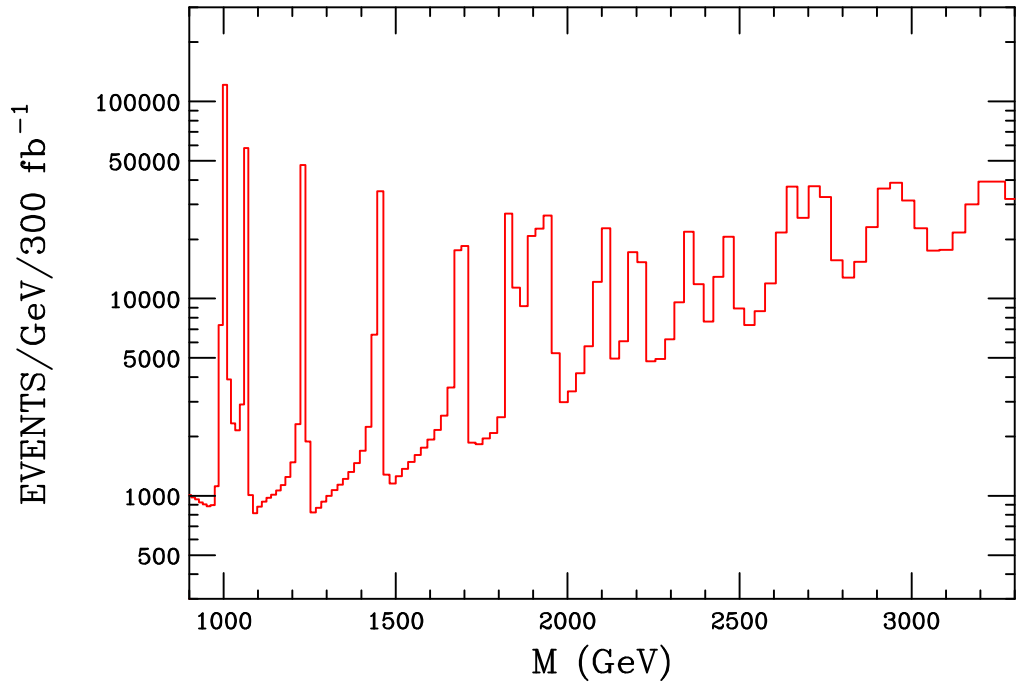}}
\vspace*{0.1cm}
\caption{Binned Drell-Yan cross section for $e^+e^-$ production at the 
LHC assuming $m_{10}=1$ TeV and all other parameters as in 
Fig.~\ref{fig1}.  The
cross section has been smeared by an electron pair mass resolution of 
$0.6\%$ as might be expected at ATLAS {\cite {atlas}}.}
\label{figLHC}
\end{figure}

Such a forest of KK
graviton resonances may also be clearly seen at the LHC. 
Fig.~\ref{figLHC}
shows a histogram of the Drell-Yan cross section into $e^+e^-$ pairs,
including detector smearing \cite{atlas} and assuming
the same model parameters as above except for $m_{10}=1$ TeV. 
The individual peaks at
smaller masses are
well separated except where they are nearly degenerate due to this choice
of model parameters. At  larger
masses there is increased overlap among the KK states and individual 
resonances may be difficult to isolate. A detailed detector study of this KK 
forest is required to determine what separation between the states is 
necessary in order to isolate the resonances.

\begin{figure}[htbp]
\centerline{
\includegraphics[width=9cm,angle=90]{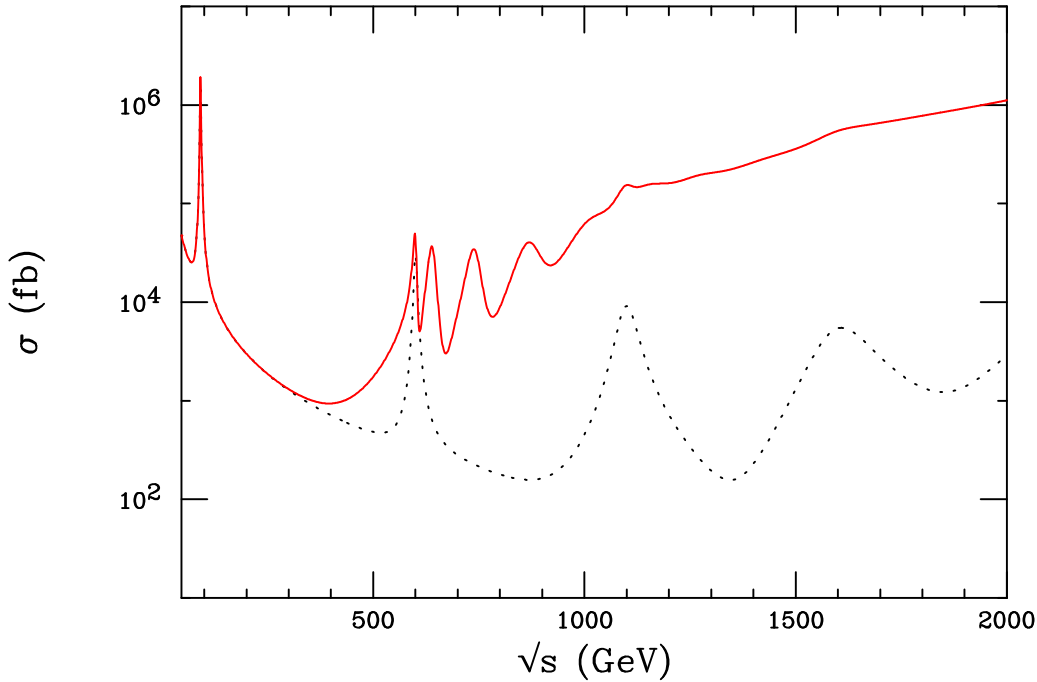}}
\vspace*{0.1cm}
\caption{The solid red curve corresponds to the cross section for 
$e^+e^- \to \mu^+\mu^-$ when the additional
dimension is orbifolded, \ie, for $S^1/Z_2$, with $m_{10}=600$ GeV, 
$k/\mpl=0.1$
and $kR=1$ being assumed. The result for the conventional RS model is 
also displayed, corresponding to the dotted curve.}
\label{fig2}
\end{figure}

We now study the variations in these results for different
choices of the model parameters. In the original RS scenario, 
the widths of the individual KK graviton resonances, as well
as the interferences between the various peaks, are determined
by  the parameter $k/\mpl$.
Increasing $k/\mpl$ leads to larger widths in the present scenario
as well, but given the rather dense forest of KK excitations, the 
individual states will now tend to overlap thus 
smearing out the cross section. To demonstrate this effect, we
consider the same
case as above but now take $k/\mpl=0.1$; the resulting spectrum
in $\epem\to\mu^+\mu^-$ is displayed in Fig.~\ref{fig2}.
Here we see that the states above $n=2,l=0$ can no longer be resolved into
individual peaks and yield a smoothly rising cross section.
Note that the state $n=1, l=4$ is barely observable. Below this
state, while the individual resonances are separated, their
shapes are distorted due to the
strong interference between the excitations. This is
particularly clear in the case of the $n=1,l=0$ state which is likely 
to be too deformed to allow for a simple extraction of the value of 
$k/\mpl$. In this case, we now find that the unitarity bound in 
Eq. (\ref{unitar}) is indeed preserved up to the
scale $\Lambda_\pi$, which is given by $\sim 1.5$ TeV for this set
of parameters.

\begin{figure}[htbp]
\centerline{
\includegraphics[width=9cm,angle=90]{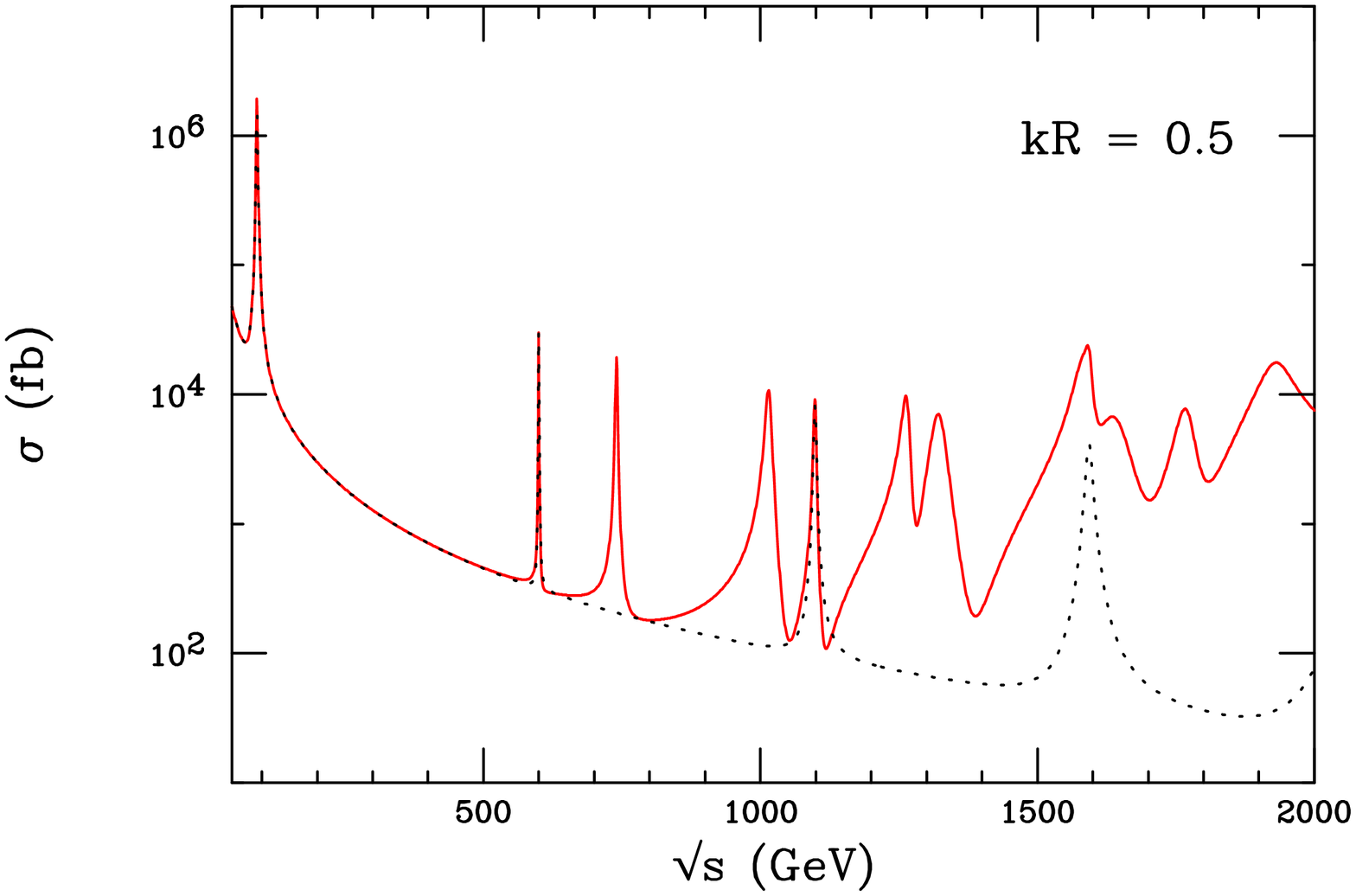}}
\vspace{5mm}
\centerline{
\includegraphics[width=9cm,angle=90]{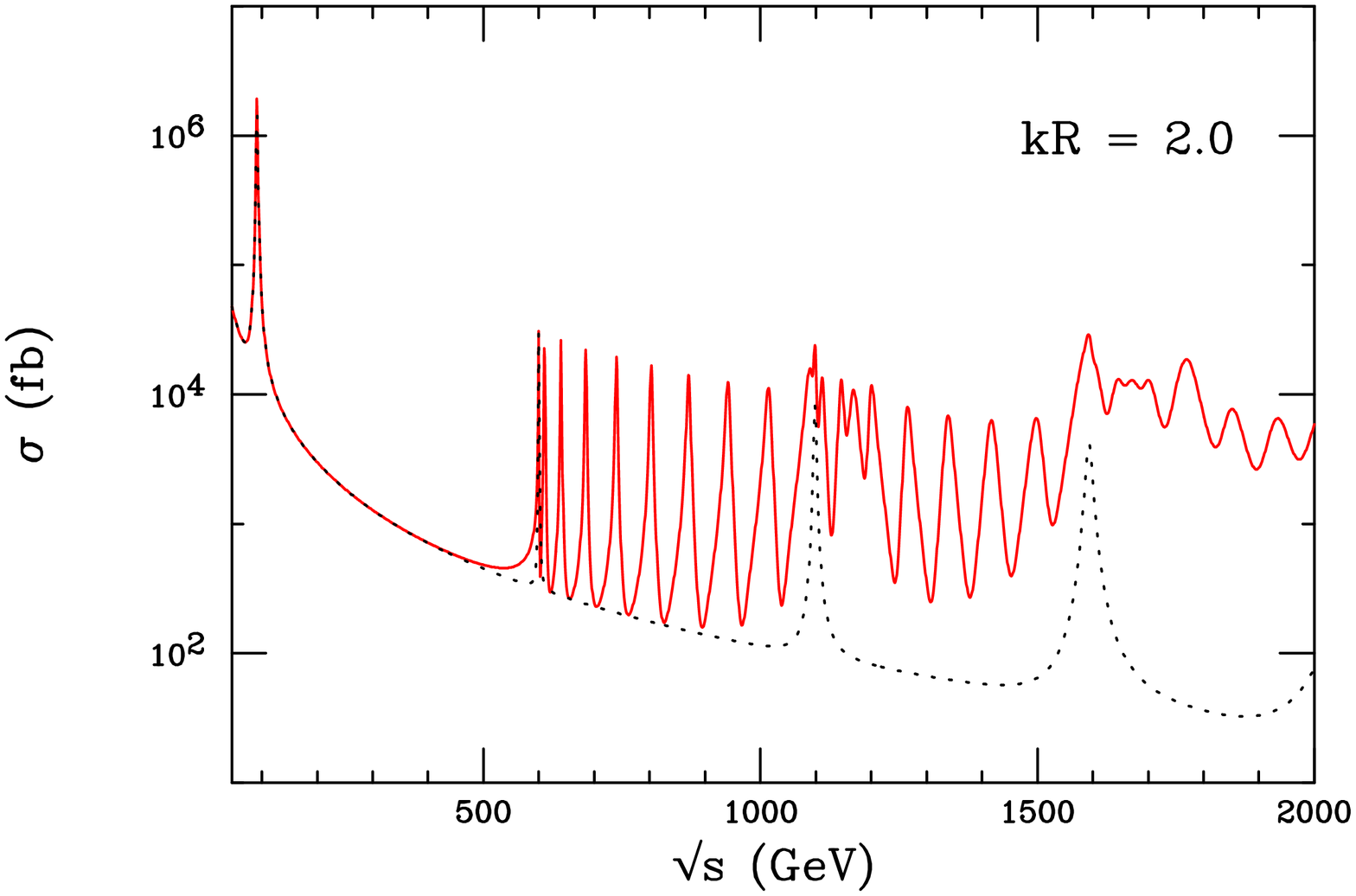}}
\vspace*{0.1cm}
\caption{The solid red curve corresponds to the cross section for 
$e^+e^- \to \mu^+\mu^-$ when the additional
dimension is orbifolded, \ie, for $S^1/Z_2$, with $m_{10}=600$ GeV, 
$k/\mpl=0.03$ and $kR=0.5$ (top), $kR=2.0$ (bottom)
being assumed. The result for the conventional RS model is 
also displayed, corresponding to the dotted curve.}
\label{fig3}
\end{figure}

Next we consider varying the values of $kR$; from Eq.(\ref{specfit})
we see that $kR$ need not be too far
away from unity in order to obtain significant modifications in the results 
shown above. Here we consider the cases $kR=0.5$ and 2 which yield
the results shown in
Fig.~\ref{fig3}. For $kR=0.5$ the states are 
more widely separated than when $kR=1$ and the individual peaks 
would be easily observed and studied in a collider detector. 
If the value of $kR$ were further reduced, the $l \neq 0$ excitations 
would quickly grow heavier, 
\eg, setting $kR=0.05(0.1)$ gives the ratio 
$m_{11}/m_{10}=5.94(3.26)$ which, for the first case, 
is significantly beyond the kinematic
range shown in the figure. If the energy range of the collider were 
restricted in comparison to such a spectrum, the $l$ excitations would go 
unobserved. As the value of
$kR$ increases, the resonance forest grows thicker and dozens of
states are seen to at least partially
overlap when $kR$ reaches a value of 2. Note that the
spacing between the $n=1,l=0$ and the $n=l=1$ state is rather small in 
this case, being only 425 MeV; this is comparable to their individual widths 
which are $\simeq 525$
MeV.  The $\mu$-pair mass resolution at a linear collider  
should be sufficient to resolve these two states if they are within
kinematic reach. However, if the value of $kR$ is increased further,
separation of these first two states
may prove difficult, especially for values of $kR$ as large
as 10 or more where the peaks would be separated by less than 100 MeV.

\begin{figure}[htbp]
\centerline{
\includegraphics[width=9cm,angle=90]{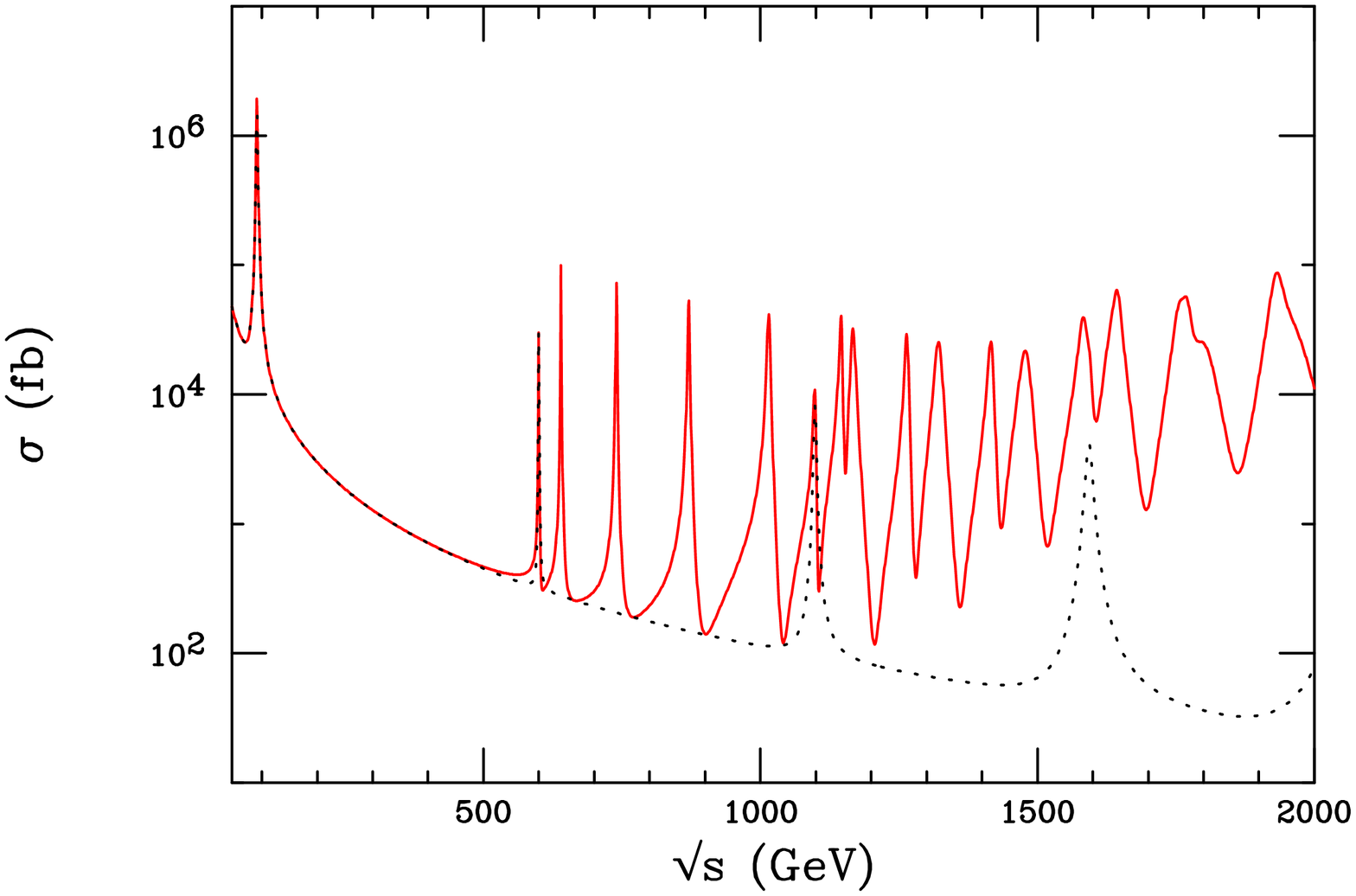}}
\vspace*{0.1cm}
\caption{The solid red curve corresponds to the cross section for 
$e^+e^- \to \mu^+\mu^-$ when the additional
dimension is not orbifolded, \ie, for the case of 
$S^1$, with $m_{10}=600$ GeV, 
$k/\mpl=0.03$ and $kR=1$.
The result for the conventional RS model is 
also displayed, corresponding to the dotted curve.}
\label{fig4}
\end{figure}

Lastly, we consider the scenario where
we do not orbifold the $S^1$. In this scenario, the
eigenfunctions and the resulting couplings are modified as discussed
above. Recall that in the case of $S^1/Z_2$,
the eigenfunctions take the form $\sim \cos l \theta$ (with $l \geq 0$)
and are normalized 
such that an additional factor of $\sqrt 2$ occurs in couplings when 
$l >0$.  This is similar to the familiar results of TeV-scale 
extra-dimensional
theories with bulk SM fields. Without orbifolding, the eigenfunctions 
are of the
form $\sim e^{il\theta}$, with $l$ of either sign, and the extra factor of
$\sqrt 2$ does not appear in the normalization. 
Thus we can treat each level with $|l| > 0$
as doubly degenerate, in contrast to the $S^1/Z_2$ case where no
degeneracy occurs. 
The results for the non-orbifolded case, taking these factors into account,
are shown in Fig.~\ref{fig4}. Except for interference
effects, we see that the $l=0$ resonances are identical in the two cases 
whereas $l\ne 0$ excitations are much more pronounced due to the
double degeneracy. Since there are essentially twice as many states, 
the overall cross section increases much more rapidly with $\sqrt s$ 
than in the orbifolded case.

\subsection{Placing the Standard Model Fields in the $S^1/Z_2$}

We now discuss the scenario where the SM fields are allowed to
propagate in the additional orbifolded $S^1/Z_2$ dimension. Note
that the orbifold symmetries must be present when the SM is in this
manifold in order to remove the extra zero-mode fields which result
from the KK expansion of the SM fields.  We remind the reader that
the placement of the SM fields in the RS bulk is problematic due
to the large contribution of the resulting KK states to precision
electroweak observables \cite{RSphen}; we thus do not consider
this option here.

\subsubsection{Constraints from Precision Measurements}

We first consider the case of placing only the SM gauge fields in
the $S^1/Z_2$ manifold.  In this scenario, the exchange of their 
KK excitations can give significant contributions to the precision
electroweak observables which leads to strong lower bounds on
the associated compactification scale \cite{big}. 
In the present case, this also yields constraints on the graviton
KK spectrum since the KK gauge and
graviton masses are linked in a very simple way. The masses of 
the KK gauge excitations are given by $m^A_l=(l/R) e^{-\pi kr_c}$.
The first gauge KK state can then be written in terms of the lowest 
lying graviton KK state as  $m^A_1=m_{10}/kRx_{10}$, implying that
$m_{10}\simeq 3.83 kR m^A_1$. Given a constraint
on $m^A_1$ from a global fit to the precision electroweak data,
this relationship places a lower bound on the first
KK graviton mass as a function of $kR$.  This in turn sets a limit on
$\Lambda_\pi$ since $m_{10}=x_{10}\Lambda_\pi k/\mpl$. The lower bound on 
$m^A_1$ in this scenario is exactly the same as that arising in the 
more familiar case of TeV-scale
flat space theories and is approximately given by 5 TeV{\cite {big}}
when the SM fermions and
Higgs fields are confined to the TeV-brane, \ie, the fixed point at 
$\theta=0$.
In this simple case this result implies that $m_{10}\geq 19 kR$ TeV and
further that
\begin{equation}
\Lambda_\pi \geq 100 kR {0.05\over {(k/\mpl)}}\rm{TeV}\,.
\end{equation}
This is uncomfortably large for $kR$ near unity and favors smaller values of 
order $\sim 0.2$ or less, which is not far from the lower end of our 
natural region.

\begin{figure}[htbp]
\centerline{
\includegraphics[width=9cm,angle=90]{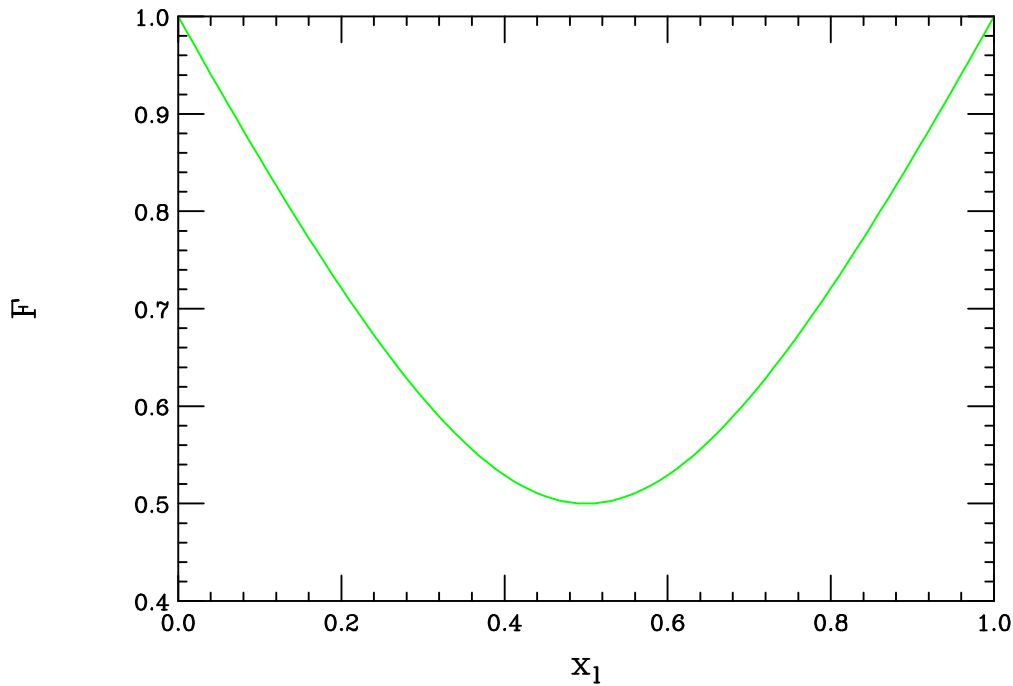}}
\vspace*{0.1cm}
\caption{Fractional softening of the $\simeq 5$ TeV lower bound on the 
lightest gauge boson KK mass as a function of the location of the 
SM leptons in the $S^1/Z_2$ manifold for the toy
model discussed in the text.}
\label{fig6}
\end{figure}

Of course, confining all the SM fields at the $\theta=0$ fixed point is 
not the scenario we envision as we want to address the problem of 
proton decay and other issues
through fermion localization{\cite {ArkSch}}.
Localization of the SM fermions in the $S^1/Z_2$ manifold will in general
lower the bound of $m^A_1\simeq 5$ TeV, since the couplings of
the SM fermions to the gauge KK states are dependent on their point of
localization. These couplings are then proportional to $\cos l\theta_f$, 
where $\theta_f$ is the localization point for a given fermion; since the 
magnitude of this quantity is always less than unity, this implies 
smaller fermionic KK gauge couplings
and thus weaker bounds from the data. An analysis of all precision 
measurements with arbitrarily localized fermions is beyond the scope 
of this paper,
but to demonstrate this phenomena we consider a toy model in which
the SM leptons are all localized at $\theta_l=\pi z_l$ and the quarks are
elsewhere. As is well known, most of the constraints set from
precision measurements arise from fits to the observables $M_W$ and 
$\sin^2 \theta_{eff}$, where the latter essentially determines 
the leptonic coupling at the $Z$-pole. These observables
depend on $z_l$ and are independent of the location of the quark 
fields and hence we can examine the
fraction by which the 5 TeV bound softens as we vary $z_l$ away
from zero. The results of this simple analysis are shown in 
Fig.~\ref{fig6}.
Here we see that within this toy model the bound can be 
softened by as much as $50\%$. 
Based on this short analysis one might expect that in a
more realistic situation with all the fermions localized in various 
places in the $S^1/Z_2$ it
may be possible to reduce this constraint even further, perhaps to the 1-2 
TeV range.  If this were so, then values of $kR$ as large as unity would 
be acceptable without having to invoke fine-tuning.

\subsubsection{Flavor Changing Graviton KK Interactions}

We now discuss other effects from localizing the SM fermions at various
points along the $S^1/Z_2$ dimension. Fermion localization may be
achieved by the use of a kink solution of a varying domain wall
scalar \cite{ArkSch}; the fermions then obtain narrow Gaussian-like 
wavefunctions with a width  much smaller than
the compactification scale for this dimension.  The $l=0$ graviton KK states
have a flat wavefunction along this dimension and hence are not
sensitive to the fermion locations. However, the wavefunction 
for the $l\neq 0$ graviton KK states goes as $\sim\cos l\theta$ and thus 
these states will have different overlaps with fermions placed at 
distinct points.  In particular, flavor changing (FC) couplings for 
these graviton KK states will result if the fermion generations are localized 
at different spots. Similar FC effects for gauge KK states have been 
studied \cite{claret} in the case where the SM fields are located
in an TeV$^{-1}$ extra dimension.  We also point out that FC graviton
KK interactions may be present \cite{gravfac} 
in conventional RS models when the SM
fermions propagate in the warped geometry.

Such FC graviton KK couplings are induced by fermion mixing.  The
overlap of the wavefunction for an $l\neq 0$ graviton KK state
with a left-handed fermion of type $i$ localized at the point 
$\theta=\theta_i$ is given by
\begin{equation}
x^i_L = \int d\theta\bar f_{L,i}(\theta)f_{L,i}(\theta)\phi^l(\theta)\,,
\end{equation}
where $f_i(\theta)$ is a gaussian of width $\sigma$, 
$\exp(-\lambda(\theta-\theta^i_{L})^2/2\sigma^2)$, and similarly for the
right-handed fermions.  Approximating the fermion gaussian wavefunctions by
a delta-function, $\delta(\theta-\theta^i)$, gives 
$x^i_{L,R}=\cos l\theta^i_{L,R}$.
The 4-dimensional interaction Lagrangian for
an $(n,l)$ graviton KK state is then
\begin{equation}
{\cal L}_{int}={-1\over\Lambda_\pi}\sum_i(\bar f^0_{L,i}\tilde T^{\mu\nu}
x^i_Lf^0_{L,i}+ \bar f^0_{R,i}\tilde T^{\mu\nu}x^i_Rf^0_{R,i})
\xi(n,l)h^{(n,l)}_{\mu\nu} \,
\end{equation}
in the fermion weak eigenstate basis, and where
we have written the stress-energy tensor as
$T^{\mu\nu}=\bar f\tilde T^{\mu\nu}f$.  In the fermion mass eigenstate
basis, this becomes
\begin{equation}
{\cal L}_{int}={-1\over\Lambda_\pi}\sum_i(\bar f_{L,i}\tilde T^{\mu\nu}
U_Lx^i_L(U_L)^\dagger f_{L,i}+ \bar f_{R,i}\tilde T^{\mu\nu}U_Rx^i_R
(U_R)^\dagger f_{R,i})\xi(n,l)h^{(n,l)}_{\mu\nu} \,,
\end{equation}
where $U_{L,R}$ represent the bi-unitary transformations which diagonalize
the fermion fields, \ie, $M^{Diag}=U_L^\dagger m^0U_R$.
We denote the mixing factors which induce FC interactions for the
graviton KK states as
\begin{equation}
X_{L,R} = U^\dagger_{L,R} \left(\begin{array}{ccc}
                    x^d & 0 & 0 \\
                     0 & x^s & 0 \\
                     0 & 0 & x^b
                    \end{array}\right) U_{L,R}\,,
\end{equation}
which gives, for example,
$X^{ij}_L=(U^\dagger_L)_{ik}x^k_L\delta_{ak}(U_L)_{aj}$.

These couplings can induce flavor changing neutral current
(FCNC) reactions which are mediated by tree-level graviton KK exchange. 
These in principle could occur at a sizeable level and pose a 
threat to this scenario.  For purposes of demonstration, 
we examine the process $B_q\to\ell^+\ell^-$ (where $q=d$ or $s$) 
in order to provide
an estimate of the magnitude of such effects.  In this case, 
using the Feynman rules of Ref. \cite{GiudiceHan},
the amplitude for the tree-level graviton KK contribution to this
decay is given by
\begin{equation}
{\cal A} = {2\over\Lambda_\pi^2}\sum_{l=1}^\infty\sum_{n=1}^\infty
\xi(n,l)^2(\bar q\tilde T^{\mu\nu}b)(\bar\ell\tilde T^{\alpha\beta}\ell)
{P_{\mu\nu\alpha\beta}\over k^2-m^2_{nl}}\,,
\end{equation}
where $P_{\mu\nu\alpha\beta}$ is the spin-sum of the graviton KK polarization 
tensors, $\xi(n,l)$ is defined in Eq. (\ref{xi}), and $k$ represents the
momentum transfer to the final state which can be neglected in comparison
to the graviton KK masses.  Here, we have suppressed the mixing factors;
their contributions will be explicitly determined below.

In order to compute the rate, we must determine the matrix element
of the hadronic current
\begin{equation}
\langle 0\, |\, \bar q\tilde T^{\mu\nu}b\, |\, B\rangle\,,
\end{equation}
where the appearance of the stress-energy tensor leads to the factor
$i[(p_b-p_q)^\mu\gamma^\nu+(p_b-p_q)^\nu\gamma^\mu]/4$,
with $p_i$ being the momenta of the $i^{th}$ quark inside the $B$ meson.
We make use of Heavy Quark Effective Theory \cite{neubert} in the
evaluation of this matrix element, which yields
\begin{equation}
{i\over 4} \langle 0\, |\, \bar q[p_B^\mu\gamma^\nu+p_B^\nu\gamma^\mu]b
\, |\, B \rangle\,,
\end{equation}
where we have neglected terms of order $\Lambda_{QCD}/m_B$.  Due to
parity considerations, only axial-vector contributions yield a non-zero 
value for this matrix element.  These are realized in the present
scenario in the case where the left- and right-handed fermions of
a given flavor are localized at separate points.  Indeed, such a
configuration could generate the flavor hierarchy by giving 
the wavefunctions of the left- and right-handed fields different degrees 
of overlap in the $S^1/Z_2$ dimension \cite{ArkSch}.
Using the familiar result
\begin{equation}
\langle 0\, |\, \bar q\gamma^\mu\gamma_5 b\, |\, B\rangle = -if_Bp_B^\mu\,,
\end{equation}
we then find
\begin{equation}
\langle 0\, |\, \bar q\tilde T^{\mu\nu}b\, |\, B\rangle={-f_B\over 2}
(X^{qb}_L-X^{qb}_R)p_B^\mu p_B^\nu\,,
\end{equation}
where $f_B$ is the decay constant of the $B$ meson and the mixing
factors are defined above. We see that this matrix element indeed
vanishes when the left- and right-handed fields are not separated.

Using these results and inserting the form of the polarization sum
\cite{GiudiceHan}, we find for the square of the amplitude
\begin{equation}
|\langle 0\, |\, {\cal A}\, |\, B\rangle|^2 = 
{f_B^2m_B^6m_\ell^2\over 2\Lambda_\pi^4} \Bigg|
\sum_{l=1}^\infty\sum_{n=1}^\infty {\xi(n,l)^2[X^{qb}_L-X^{qb}_R]
[x^\ell_L-x^\ell_R]\over m^2_{nl}} \Bigg|^2 \,,
\end{equation}
which leads to the partial decay width for this contribution
\begin{equation}
\Gamma(B_q\to\ell^+\ell^-)_{grav} 
= {f_B^2m_B^5m_\ell^2\over 32\pi\Lambda^4_\pi
m^4_{10}}\bigg[ 1-{4m^2_\ell\over m_B^2}\bigg]^{1/2}\Bigg| 
\sum_{l=1}^\infty\sum_{n=1}^\infty{ \xi(n,l)^2x_{10}^2
[X^{qb}_L-X^{qb}_R][x^\ell_L-x^\ell_R]\over x^2_{nl}}\Bigg|^2\,.
\end{equation}
The overall factor of $m_\ell^2$ arises from helicity suppression.
Here, $x_{nl}$ are the roots which determine the graviton KK masses as
discussed above.  
A numerical evaluation of this width for the case of $B_s\to\mu^+\mu^-$,
with $m_{10}=600$ GeV and $k/\mpl=0.1$, yields the result
\begin{equation}
\Gamma(B_s\to\mu^+\mu^-)_{grav}= (2.83\times 10^{-23}{\rm MeV})
\Bigg| 
\sum_{l=1}^\infty\sum_{n=1}^\infty{\xi(n,l)^2x_{10}^2
[X^{qb}_L-X^{qb}_R][x^\ell_L-x^\ell_R]\over x^2_{nl}}\Bigg|^2\,.
\end{equation}
In order to estimate the size of the mixing factors, we assume
$x^s_{L,R}=x^d_{L,R}$ and make use of
unitarity which yields
\begin{equation}
X^{sb}_{L,R}=(x^b_{L,R}-x^s_{L,R})(U^\dagger_{L,R})_{sb}(U_{L,R})_{bb}\,.
\end{equation}
Taking the most optimistic case with
the left-(right-)handed fermions being located at $\theta_l=0\,(\pi)$,
and setting $(U^\dagger_{L,R})_{sb}(U_{L,R})_{bb}=\lambda^3$ where
$\lambda$ represents the Cabbibo mixing angle, we find the graviton
KK contribution to the branching ratio for this decay to be
\begin{equation}
BR(B_s\to\mu^+\mu^-)_{grav}\sim 10^{-16}
\end{equation}
at the most.  Compared to the SM branching fraction of $\sim 10^{-9}$,
we see that the KK graviton contributions are safely below the SM
prediction.  We thus conclude that low-energy FCNC do not pose a
dangerous threat to this scenario.

In addition, these flavor changing KK graviton couplings may be observed in
the decays of the $n\geq 3$ graviton states at colliders, 
\ie, $G^{(nl)}\to t\bar c$.  Such FC graviton decays have been discussed
in \cite{gravfac}.

\section{The Graviton KK Spectrum and Couplings: Curved Manifolds}

Here, we find that for $S^\delta$ with $\delta > 1$ a new feature arises
due to the positive curvature of the $S^\delta$.  As we will see below,
this curvature contributes the same way as a negative
cosmological constant to the effective 5-d RS geometry.  Thus,
it is possible to set the cosmological constant in the
non-spherical dimensions to zero and {\it still generate a warped geometry}.  
This fixes the relation
between $k$ and $R$ to a value which is within our natural range, 
thus reducing the number of free parameters,
and yields a warp factor that
is set geometrically by the radius of the $S^\delta$.

To demonstrate the effects of higher dimensional
spheres, we will present the relevant formulae for the case
of $S^2$.  The geometries with $\delta > 2$ are generalizations
of the $\delta = 2$ scenario and we briefly comment on them at the
end of this section. A simple curved manifold such as $S^\delta$
could in principle address proton decay and other model building issues,
however the mechanism for localizing fermions on a curved manifold
has not yet been demonstrated.

\subsection{Formalism}

For $\delta = 2$, we parameterize the metric as
\begin{equation}
ds^2 = e^{-2\sigma} \eta_{\mu \nu} dx^\mu dx^\nu + r_c^2 \,
d\phi^2 + R^2 (d\theta^2 + \sin^2\theta \, d\omega^2),
\label{s2met}
\end{equation}
where we now have $\theta \in [0, \pi]$ and
$\omega \in [0, 2 \pi]$.  The cosmological constant and
energy-momentum tensors now have the general forms
\begin{equation}
\Lambda^A_B =
{\rm diag}(\Lambda, \Lambda, \Lambda, \Lambda, \Lambda,
\Lambda_\theta, \Lambda_\omega),
\label{cctensor2}
\end{equation}
and
\begin{equation}
T^M_N = - \left\{\delta(\phi)\left(\begin{array}{cccc}
V^h \delta^\mu_\nu & 0 & 0 & 0 \cr
0 & 0 & 0 & 0 \cr
0 & 0 & V_\theta^h & 0 \cr
0 & 0 & 0 & V_\omega^h
\end{array}\right) + \delta (\phi - \pi)
\left(\begin{array}{cccc}
V^v \delta^\mu_\nu & 0 & 0 & 0 \cr
0 & 0 & 0 & 0 \cr
0 & 0 & V_\theta^v & 0 \cr
0 & 0 & 0 & V_\omega^v
\end{array}\right)\right\},
\label{emtensor2}
\end{equation}
respectively.

As before, using the metric in (\ref{s2met}) and
solving Einstein's equations component by
component, will fix the various parameters of the model.
For example, the $(4, 4)$ component yields the following
for the warping scale
\begin{equation}
k^2 = \frac{-\Lambda}{24 M_F^5} + \frac{1}{6 R^2}.
\label{s2k}
\end{equation}
Here, we explicitly see that the curvature of the sphere contributes
to the scale $k$ in the same fashion as a negative cosmological
constant.  The other non-trivial
components of Einstein's equation yield the following
relations
\begin{equation}
V^h = -V^v = 24 \, M_F^5 \, k  \, \, \, ; \, \, \,
V_\omega^i = V_\theta^i = \frac{4}{3} V^i,
\label{s2tensions}
\end{equation}
where $i = h, v$ and
\begin{equation}
\Lambda_\omega = \Lambda_\theta = \frac{5 \Lambda}{3} -
\frac{20}{3 R^2}M_F^5.
\label{s2ccs}
\end{equation}
Eqs.(\ref{s2k}), (\ref{s2tensions}), and (\ref{s2ccs})
show that we now have the freedom to choose $\Lambda$
to be zero, or even positive, as long as $k^2 > 0$.  In
particular, for $\Lambda = 0$ we obtain uniquely the relation 
$k = 1/(\sqrt {6} R)$ which lies in our natural range of values for $kR$.
In this case, the warped geometry in the RS model arises {\it solely} 
from the curvature of the $S^2$ manifold.

The relation between $\mpl$ and $M_F$ is now given by
\begin{equation}
\mpl^2 = \frac{4 \pi R^2}{k} \, M_F^5[1 - e^{-2\sigma(\pi)}].
\label{MPrel2}
\end{equation}
We again focus on metric perturbations of the form in (\ref{metpert}).
Here, due to the spherical symmetry of $S^2$,
we choose the following  KK expansion for the graviton
\begin{equation}
h_{\mu\nu}(x, \phi, \theta, \omega) = \sum_{n, l, m}
h^{(n, \, l, \, m)}_{\mu\nu}(x)
\frac{\chi^{(n,\, l)}(\phi)}{\sqrt {r_c}}
\frac{Y^m_l(\theta, \omega)}{R},
\label{gravKKs2}
\end{equation}
where the $Y^m_l(\theta, \omega)$ are the spherical harmonics.
The above expansion yields an equation of motion which is
similar to that obtained for the case with $\delta = 1$.  We have
\begin{equation}
-\frac{1}{r_c^2} \frac{d}{d\phi}\left(e^{-4 \sigma}
\frac{d}{d\phi} \chi^{(n,\, l)}(\phi)\right) +
e^{-4 \sigma}\, \frac{l(l + 1)}{R^2} \, \chi^{(n,\, l)}(\phi)
= e^{-2 \sigma} m_{nl}^2 \chi^{(n,\, l)}(\phi).
\label{KKeqs2}
\end{equation}
The solution for $\chi^{(n,\, l)}$ is given in (\ref{chi}), where
we now have
\begin{equation}
\nu \equiv \sqrt {4 + \frac{l(l + 1)}{(k R)^2}}\, .
\label{nus2}
\end{equation}
Substituting the definition (\ref{nus2}) for $\nu$ in
Eqs.(\ref{Nq}) and (\ref{xq}) we obtain the
normalization coefficients $N_{nl}$ and the roots $x_{nl}$
which determine the masses $m_{nl}$ as before.

We are now in a position to derive the coupling of
the graviton KK tower to the 4-d localized fields.  By
spherical symmetry we may choose any point on the sphere
to place the 4-d fields.  A particularly convenient
choice is $\theta = 0$, for which we have
\begin{equation}
Y^m_l (0, \omega) = \sqrt{\frac{2 l + 1}{4 \pi}} 
\delta_{m, 0} \,\,.
\label{ykl}
\end{equation}
This demonstrates that, at this point, the coupling is independent
of $\omega$ and that for any $l$ only the $m = 0$ states couple.
For the case when the 4-d fields are localized at $\theta=\pi$,
the above is modified by the overall factor $(-1)^l$.

The coupling of the 7-d graviton to the 4-d energy-momentum
tensor is given by
\begin{equation}
{\cal L} = -\frac{1}{M_F^{5/2}} h^{\mu\nu}(x, \pi, 0, \omega)
T_{\mu\nu}(x).
\label{7dcoup}
\end{equation}
Using the solutions to (\ref{KKeqs2}),
Eq.(\ref{ykl}), and substituting the KK expansion
(\ref{gravKKs2}) in the above, we obtain
\begin{equation}
{\cal L} = -\frac{1}{\mpl} h_{\mu\nu}^{(0, \, 0, \, 0)}(x)
T^{\mu\nu}(x)
-\frac{1}{\Lambda_\pi} T^{\mu\nu}(x)
\sum_{l = 0}^\infty \sum_{n = 1}^\infty \eta(n, l)
h_{\mu\nu}^{(n, \, l, \, 0)}(x),
\label{4dcoups2}
\end{equation}
where
\begin{equation}
\eta (n, l) = \sqrt{2 l + 1}\left[1 - \frac{l(l + 1)}
{(k R x_{nl})^2}\right]^{-1/2}.
\label{eta}
\end{equation}

As in the case of the $S^1$ and $S^1/Z_2$ manifolds, we see that
only warped graviton KK states exist (\ie, there are no $\chi^{(n,l)}=$
constant modes for $l\neq 0$).

\subsection{Numerical Results}

\begin{figure}[htbp]
\centerline{
\includegraphics[width=9cm,angle=90]{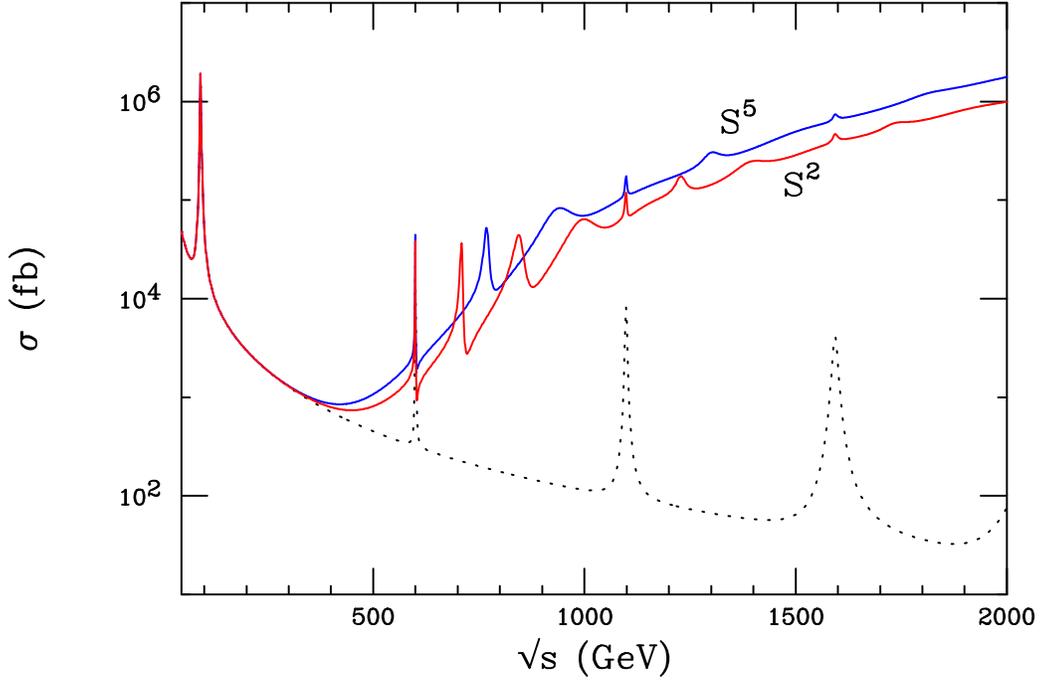}}
\vspace*{0.1cm}
\caption{The cross section for 
$e^+e^- \to \mu^+\mu^-$ when the additional
dimension is curved with $m_{10}=600$ GeV, 
$k/\mpl=0.03$ and $kR=1$.  The lower solid (red) curve corresponds
to the manifold $S^2$, while the upper solid (blue) curve represents
the case $S^5$.
The result for the conventional RS model is 
also displayed, corresponding to the dotted curve.}
\label{fig5}
\end{figure}

We first display the resulting KK spectrum from this analysis for
the case of $S^2$ in Fig.~\ref{fig5}. Compared to the case with
$\delta=1$, we now have more KK states, however the number of states
which couple to the 4-d fields is the same; the strength
of this coupling is now stronger.  From the figure, we see that
the first few resonances are clearly observable, 
whereas the higher resonances are blurred.  In this case, we see
that the sharp rise in the cross section results in an early onset
of unitarity violation, and hence the model parameters in this case
are restricted.

As a last possibility, we briefly discuss the extension of $S^2$ to
the case $S^p$, where $2< p \leq 5$.  Although we have not
explicitly performed a detailed derivation of the resulting KK
expansion and couplings to the 4-d fields in this higher dimensional case, 
we can make use of the work of {\cite{psphere}} to anticipate
how this generalization might proceed. We consider the case 
where all SM fields
are constrained to the fixed point,
which corresponds to $\theta=0$ for
arbitrary $p$. (In addition to $\theta$ there will also be $p-1$ 
azimuthal-like co-ordinates $\phi_i$.)
We only consider how the masses, couplings and degeneracies of the 
KK states might be altered for arbitrary $p$.  Comparing to the case
of $S^1$, the results of \cite{psphere} suggest
that the terms proportional to $l^2$ in both the 
expressions for $\nu$ and
$\xi$ are given by the eigenvalues
of the square of the angular momentum operator. Generalizing to
a $p$-sphere, we would have $l^2 \to L^2\to l(l+p-1)$. 
We assume that the KK excitations with non-trivial azimuthal quantum
numbers will not couple to the SM fields at the fixed point. 
This implies that although each $l$ level is
multiply degenerate, only one of these graviton KK states will couple 
to the SM fields. In addition, we assume that 
the overall couplings pick up an
additional factor due to the normalization of the angular wavefunctions
which can be determined for any value of $p$ {\cite {psphere}}.
We note that each of these modifications are verified in the case
of $S^2$ by our preceding analysis.

Fig.~\ref{fig5} also displays our results for the KK spectrum in
the case of an $S^5$ manifold,
making use of our assumptions listed above regarding the 
generalization to higher dimensional spheres.  With the above
assumed modifications, 
we see that the individual KK states become somewhat
less distinct in higher dimensional spherical geometries.
Note also that the mass of the $n=l=1$ resonance tends to increase as the
dimensionality of the sphere grows.

\section{Conclusions}

The RS model offers a natural 5-d mechanism for generating the gauge 
hierarchy.  However, it leaves a number of questions, such as proton 
decay, flavor, and the nature of quantum gravity, unanswered.  Whether 
from a model building point of view, or from a more fundamental standpoint, 
it may be necessary to embed the RS scenario in a higher dimensional space 
to address such questions.

In this paper, we assumed that the RS model is endowed with $\delta \geq 1$ 
additional dimensions compactified on a manifold ${\cal M}^\delta$.  The 
background geometry was taken to be of the form $|AdS_5| \times 
{\cal M}^\delta$, a direct product of the original RS geometry and 
${\cal M}^\delta$.  We considered two classes of manifolds: flat  
and curved.  We then studied the parameters required to establish these 
backgrounds, as well as the resulting graviton KK spectrum, couplings, 
and the corresponding collider phenomenology.
For simplicity, we chose ${\cal M}^\delta = S^\delta$.  In the case of flat 
geometries, we studied $S^1/Z_2$ and $S^1$ as 
simple representative manifolds.  
The $S^1/Z_2$ case is of particular interest, since the fermion localization 
mechanism on this space has been studied in detail \cite{ArkSch} 
and can address 
questions of proton stability and flavor, in a simple geometric way.  We 
analyze the collider spectroscopy of this model and find that a forest of new 
KK graviton states, in addition to the original RS modes, appear at the weak 
scale.  This is a generic signature of the models that we study.  
The new modes originate from the `angular' graviton excitations 
over $S^\delta$.

The size $R$ of the radius of the $S^1$, in units of the inverse 5-d curvature 
scale $k$, is of key importance and sets the scales of
mass and separation of the KK modes.  We also find that for $k R \sim 1$, the 
couplings of different light KK modes to the 4-d SM fields are measurably 
non-universal.  This is in contrast to the 5-d RS model where such 
non-universality was exponentially suppressed.  If the SM fields reside 
in the $S^1/Z_2$ extra dimension, the couplings of various localized fermions 
to a particular KK state of the graviton or gauge fields would be 
non-universal, and we expect tree-level FCNC effects to arise in KK 
mediated processes.  We have checked that the size of such FCNC effects
do not occur at a dangerous level in this scenario.  We also show 
that over the natural range of parameters in this model, observable
experimental signatures can be expected at the LHC or a future $e^+ e^-$ 
collider.  

In the case of curved manifolds, we studied $S^2$ as an example.  Here, a new 
feature arises which is the possibility of using the positive curvature of 
$S^2$ in order to generate a negative 5-d cosmological constant $\Lambda$.  
This allows us to choose $\Lambda = 0$ along the original 5-d spacetime of 
the RS model and generate the required warping of $AdS_5$ from the curvature 
of the $S^2$.  This fixes all the parameters of the model in terms of only 
three quantities: the fundamental scale $M_F$, the size of the $AdS_5$ slice 
$r_c$, and the radius $R$ of $S^2$.  We note that positive values of 5-d 
$\Lambda$ are also allowed, as long as $k^2 > 0$.  
In going from $S^1$ to $S^2$ the collider phenomenology does not change 
significantly.  However, the
angular KK states get more strongly coupled with growing
mass, resulting in wider KK resonances with more overlap.  This has the 
effect of smearing the spectrum.  

Although we only studied $S^1/Z_2$, $S^1$, and $S^2$, generalization to 
$S^\delta$, $\delta > 2$, is straightforward.  We comment on the expected 
behavior, using $S^5$ as an example, without a detailed analysis.  A simple 
modification of our results for $S^2$ suggests that compactifying 
on $S^5$ does not change the collider phenomenology significantly.  Future 
directions for expanding our work include incorporating the effects of KK 
self-couplings as well as gravi-scalar and gravi-vector interactions 
(present for $d > 5$), and investigating the possibility of 
obtaining new warped 5-d effective theories from different 
choices of ${\cal M}^\delta$.

\noindent{\bf Acknowledgements}

The authors would like to thank Tim Barklow, Tony Gherghetta, Juan Maldacena,
and Massimo Porrati for discussions related to this work. The work of
H.D. was supported by the US Department of Energy under contract
DE-FG02-90ER40542.

%
\def\MPL #1 #2 #3 {Mod. Phys. Lett. {\bf#1},\ #2 (#3)}
\def\NPB #1 #2 #3 {Nucl. Phys. {\bf#1},\ #2 (#3)}
\def\PLB #1 #2 #3 {Phys. Lett. {\bf#1},\ #2 (#3)}
\def\PR #1 #2 #3 {Phys. Rep. {\bf#1},\ #2 (#3)}
\def\PRD #1 #2 #3 {Phys. Rev. {\bf#1},\ #2 (#3)}
\def\PRL #1 #2 #3 {Phys. Rev. Lett. {\bf#1},\ #2 (#3)}
\def\RMP #1 #2 #3 {Rev. Mod. Phys. {\bf#1},\ #2 (#3)}
\def\NIM #1 #2 #3 {Nuc. Inst. Meth. {\bf#1},\ #2 (#3)}
\def\ZPC #1 #2 #3 {Z. Phys. {\bf#1},\ #2 (#3)}
\def\EJPC #1 #2 #3 {E. Phys. J. {\bf#1},\ #2 (#3)}
\def\IJMP #1 #2 #3 {Int. J. Mod. Phys. {\bf#1},\ #2 (#3)}
\def\JHEP #1 #2 #3 {J. High En. Phys. {\bf#1},\ #2 (#3)}

\end{document}